\documentclass[reprint,amsmath,amssymb,aps]{revtex4-1}
\usepackage{graphicx}
\usepackage{tikz}
\usepackage{bm}
\usepackage{dcolumn}
\usepackage{comment}
\usepackage{lineno,blindtext}
\usepackage{amssymb}
\newcommand{\beq}{\begin{equation}}
\newcommand{\eeq}{\end{equation}}
\newcommand{\ber}{\begin{eqnarray}}
\newcommand{\eer}{\end{eqnarray}}
\newcounter{saveeqn}%
\newcommand{\alpheqn}{\setcounter{saveeqn}{\value{equation}}%
\stepcounter{saveeqn}\setcounter{equation}{0}%
\renewcommand{\theequation}
             {\mbox{\arabic{saveeqn}-\alph{equation}}}}%
\newcommand{\reseteqn}{\setcounter{equation}{\value{saveeqn}}%
\renewcommand{\theequation}{\arabic{equation}}}%
\begin{document}

\title{Persistence Exponent  for  the Simple Diffusion Equation: The Exact
Solution for any  Integer Dimension}

\author{Devashish Sanyal}
\email{deva\_sans@yahoo.co.in}
\affiliation{ Theoretical Condensed Matter \\
Institute of Physics \\
 Bhubaneswar 751005, INDIA \\
 }
\date{\today}
\begin{abstract}

The persistence exponent $\theta_o$  for the simple   diffusion  equation
 ${\phi}_t({\it x},t) = \triangle \phi (x,t)$  , with random Gaussian initial condition {\color{red},}
 has
 been calculated  exactly  using a method known as selective averaging. The
 probability that the value of the field $\phi$ at a specified  spatial
coordinate remains positive  throughout  for  a certain time $t$  behaves as
 $t^{-\theta_o}$ for asymptotically large time $t$. The value of $\theta_o$,
  calculated here   for any integer dimension $d$,  is $\theta_o = \frac{d}{4}$
 for $d\leq 4$ and $1$ otherwise. This
 exact theoretical result is being  reported possibly for  the first time  and
is not in agreement
 with  the accepted values $ \theta_o = 0.12, 0.18,0.23$ for $d=1,2,3$
 respectively.
\end{abstract}
\maketitle
\section{Introduction}
    The problem  in the present paper is to find the persistence exponent
 for the simple diffusion equation ${\phi}_t({\it x},t) = \triangle \phi (x,t)$.
 The diffusion equation is an equation that has no stochasticity. In the
 present problem the stochasticity is introduced through the random initial
 conditions. The problem is about evaluating  the probability of a certain
 event. The event is  $\phi$ at a
 specified location remains positive throughout the time evolution till
  a certain time  $t$ i.e. the $\phi$ at the location does not change sign
 even once. This probability for asymptotically large time is characterised
 by an exponent $\theta_o$ called the persistence exponent.
Persistence exponent  for the diffusion equation has been a
 subject of   interest to physicists
\cite{majumdar1996nontrivial,
derrida1996persistent,ehrhardt2002series,
 hilhorst2000persistence, schehr2008real,
 poplavskyi2018exact, barbier2020anomalous} etc ,
 researchers in mathematics and statistics 
\cite{aurzada2015persistence},\cite{dembo2015no} etc
 as well as experimentalists \cite{wong2001measurement}.
The interest in the persistence exponent
 is  just not confined to the diffusion equation but to other areas of
 non-equilibrium physics. Among them random walk \cite{schwarz2001first}, walk in a random
 environment with or without bias \cite{le1999random}, surface growth
\cite{krug1997persistence} , diffusing
 particle in a random potential with a small concentration of absorbers
\cite{le2009sinai}, behaviour of financial markets 
\cite{constantin2005volatility}  etc  are worth mentioning.
There are  few exact calculations for the persistence exponent
 in the literature. The case of a simple random
 walk in one dimension
 gives the exponent $\theta_o = \frac{1}{2}$. Even the calculation of
 persistence exponents for Gaussian processes may not be straight forward.
  \\

  We revisit the problem of simple diffusion . It is   strongly non-Markovian
 in nature. The problem involves the partial differential equation
${\phi}_t=\triangle\phi$ with random Gaussian initial conditions.
 It appears to remain an unsolved problem  even though  results \cite{
majumdar1996nontrivial,derrida1996persistent} and
several others  have been
 reported.  The problem of diffusion may  require a better understanding
 in the context of  persistence. The  article tries to find an
 exact solution to the problem.\\
\section{Simple Diffusion Equation , Random Initial Conditions and Persistence
 Exponent}

  The  diffusion equation ${\phi}_t=\triangle\phi$ is  a coarse
 grained  differential equation  whose solution is uniquely
 determined by the initial condition. In the present problem, the initial
 condition is not fixed but is chosen from a distribution. The initial
 value of $\phi$ at every coordinate is chosen from a Gaussian distribution
  with mean $0$ , variance $k$ and the initial values of $\phi$ at any two coordinates are statistically
 independent.

  In order to
calculate persistence
 exponent $\theta_o$  we have to calculate the probability that  the
 field $\phi$ at a specified coordinate does not flip sign even once throughout
 a time $t$. This probability $\mathcal{P}^{+}(t)$
of $\phi$ always remaining +ve behaves in the
 limit of asymptotically large  time as $\mathcal{P}^{+}(t)\sim t^{-\theta_o}$. This is
 true for a non-stationary process like in the present case. In 
 this article any position $x$ coordinate is a vector quantity in a $d$ dimensional
 space
 The moments of the initial condition distribution described  above  are are given by

\alpheqn
\beq\label{first}
\langle \phi (x,0)\rangle = 0
\eeq

\beq\label{second}
\langle \phi (x_1,0) \phi (x_2,0)\rangle = k{{\delta}^{(d)}}(x_1 - x_2)
\eeq
\reseteqn

 where k is the variance ofthe distribution. The solution  for the diffusion equation may
 be written in terms of the initial condition as
\beq\label{third}
\phi (x,t) = \int \,{d^d}{x}^{\prime} G(x -{x}^{\prime} , t)
\phi ({x}^{\prime} ,0)
\eeq
 where $G(x,t) = {(4\pi t)}^{-d/2} \exp({-x^2}/{4t}) $. The plan
 for the evaluation of the exponent is as follows. First, we have to
 calculate the probability  of  $\phi$ attaining a specific final value
 $\beta$ at a certain $x=x_o$  starting from a definite  initial value
 $\alpha$
 of $\phi$   at $x=x_o$. In order to evaluate it we use  the
 method of selective averaging.  The paths  that take the initial $\alpha$
 to the final value $\beta$ also comprise those   where $\phi(x_o)$ flips
 sign atleast once  during time evolution. The probability of such paths  is 
 to be subtracted out. Finally, there has to be an integration over the final
 $\beta$ from $0$ to $\infty$, followed by an integration over $\alpha$
 from $0$ to $\infty$. \\
 Selective averaging means
 averaging over the initial field
 $\phi (x,0)$, except when $x=x_o$. In other words, the averaging is done  over all  the initial
 configurations  such that $\phi$ at $x=x_o$is kept fixed at $\alpha$(say) i.e
 $\phi (x_o,0)=\alpha$ while for $x\neq x_o$ $\phi$ varies according to
 Gaussian distribution. In this paper  the selective  distribution, denoted
by  subscript $s$, is
 characterized by the moments,
\alpheqn
\beq\label{fourth}
{\langle\phi (x,0)\rangle}_s = \alpha\delta^{(d)} (x-x_o)
\eeq
\beq\label{fifth}
{\langle\phi (x_1 , 0)\phi (x_2 ,0)\rangle}_s = \big\{k + [{\alpha}^{2}
-k]\delta^{(d)} (x_1- x_o)\big\}\delta^{(d)} (x_1 - x_2)
\eeq
\reseteqn

 It may  be verified from (\ref{fourth}), (\ref{fifth}) that if 
$x\neq x_o$, $x_1\neq x_o$,
 $x_2\neq x_o$, we get
 (\ref{first}),(\ref{second})
and for   $x=x_1=x_2=x_o$, (\ref{fourth}),(\ref{fifth}) give $\alpha$, $\alpha^{2}$ as expected. Using (\ref{fourth}),(\ref{fifth}), we can calculate the moments of the random variable
 $\phi (x_o,t)$,
\beq\label{sixth}
{\langle\phi (x_o,t)\rangle}_s = {(4\pi t)}^{-d/2}\alpha
\eeq

\ber\label{seventh}
\lefteqn{{\langle{\phi}^{2}(x_o,t)\rangle}_s =} \nonumber\\
&& \int\,d^{d}{x_1}^{\prime}\,d^{d}{x_2}^{\prime}{(4\pi t)}^{-d}
\exp [-\frac{{(x_o -{x_1}^{\prime})}^{2}}{4t}]\times \nonumber\\
&&\exp [-\frac{{(x_o -{x_2}^{\prime})}^{2}}{4t}]
{\langle\phi ({x_1}^{\prime},0)\phi ({x_2}^{\prime},0)\rangle}_s \nonumber\\
&&   =k \int\,d^{d}{x_1}^{\prime}{(4\pi t)}^{-d}
\exp [-\frac{{(x_o -{x_1}^{\prime})}^{2}}{2t}] \nonumber\\
&&- \frac{k}{{(4\pi t)}^{d}}  + \frac{{\alpha}^{2}}{{(4\pi t)}^{d}}
\eer

 While evaluating the second order moment, we have used the relation in
(\ref{fifth}).
 Hence the mean and the variance of the distribution for $\phi(x_o, t)$,
 represented by $\mu$ and
 ${\sigma}^{2}$ respectively, are
\alpheqn
\beq\label{eight}
\mu ={\langle\phi (x_o,t)\rangle}_s = {(4\pi t)}^{-d/2}\alpha
\eeq
\ber\label{ninth}
\lefteqn{{\sigma}^{2} =
{\langle{\phi}^{2}(x_o,t)\rangle}_{s} - {{\langle\phi(x_o,t)
\rangle}_{s}}^{2}} \nonumber\\
&& = k {(4\pi)}^{-d} 2^{(d/2 -1)}k_d \Gamma (d/2) t^{-d/2} -
 k{(4\pi t)}^{-d}
\eer
\reseteqn
 In the above
 equation $k_d$ denotes
 the angular integration in $d$ dimensional space while $\Gamma$ represents
 the usual Gamma function. It may be mentioned that $\phi (x,t)$
 in (\ref{third}) is Gaussian
 irrespective of whether $\phi (x^{\prime},0)$  , the initial Gaussian field,
 is correlated or  not.   In the present
 case,though, the initial field is uncorrelated and $\phi(x,t)$ can
 be proved to be Gaussian  using characteristic functions
 in probability theory \cite{mathews1970mathematical}. It may be noted
 that the $\delta$ function
 distribution is  the limiting case of a Gaussian distribution.  The expression for the
  conditional probability for starting at
  $\alpha$ and  being between $ \beta$  and $\beta + d\beta$
at time $t_1$ is
\beq\label{tenth}
P(\beta |\alpha)d\beta = \frac{1}{\sqrt{2\pi}\sigma} \exp\big[\frac
{-{(\beta - \mu)}^{2}}{2{\sigma}^{2}}\big]d\beta
\eeq
 where $\mu =\mu(\alpha,t_1)$ and
 $\sigma = \sigma(t_1)$. This probability   considers
 all the paths that start from $\alpha$ to be between $\beta$
and $\beta +d\beta$ at  time
$t_1$ including ones that flip en route $\beta$ as depicted in Fig 1. Fig 1 is
 the projection of the trajectory of the system in the infinite dimensional
$\Phi - t$ space on to the $\phi(x_o) - t $ plane.

\tikzstyle{place}=[circle,draw=black!50,fill=black!50,inner sep=0pt,minimum
 size=2mm]
\begin{center}
\begin{tikzpicture}
\draw[->] (-3,0) -- (3,0)
node[below right] {$t$};
\draw[->] (0,-3) -- (0,3)
node[left] {$\phi(x_o)$};
\node at (0,1) [place] {};
\node at (0,-1) [place] {} ;
\node at (2,2) [place] {};
\draw (0,1) parabola (2,2);
\path[draw=blue] (0,1) parabola (0.8,-2);
\path[draw=blue] (0.8,-2) parabola (2,2);
\draw (0,-1) parabola (2,2);
\draw (2.3,2.3) node{B};
\draw (-0.5,1) node{A};
\draw (-0.5,-1) node{$\bar{A}$};
\draw (0.6,0.3) node{D};
\draw (2,-0.2) node{$t_1$};
\draw (0,-3.5) node{\bf{Fig 1:} {\small\texttt
{Projection of $\Phi\ - t$ trajectory }}};
\draw (0,-3.8) node{{\small\texttt{onto the $\phi (x_0) - t$ plane}}};
\begin{scope}[very thick,dashed]
\draw (2,2) -- (2,0);
\end{scope}
\end{tikzpicture}
\end{center}

$A(0,\alpha)$ represents the starting point and $B(t_1,\beta)$, the destination.
 $AB$ represents
 a path along which $\phi(x_o)$ does not flip and $ADB$ (blue curve)
 is  a typical path along
 which $\phi(x_o)$ flips. Such paths have to be excluded.
 The probability of reaching from $A$ to the neighborhood $B$
 at asymptotically large time $t_1$ without flipping is
 given  by,

\beq\label{impequation}
P^{+}\big(\beta|\alpha\big)d\beta=
P\big(\beta|\alpha\big)d\beta- P\big(\beta|-\alpha\big)\big( 1 +O(t^{-1})\big)
d\beta
\eeq

 The second term represents the probability of paths such as $\bar{A}DB$
 originating from $\bar{A}(0,-\alpha)$
 and terminating in the neighborhood of  $B$ at $t_1$. (\ref{impequation})
  is not be confused with  the method of images in
\cite{chadrasekhar1989stochastic}. (\ref{impequation})   follows a very different
 logic in the present case and holds good asymptotically. To prove
(\ref{impequation})
 we will show that  there is a one to one mapping from a path $A\rightarrow
 B$ to a path $\bar{A}\rightarrow B$
 and that the probability of two such paths converge asymptotically.
 This part is explained in  $(i)$ in  what follows. Further, to
 justify (\ref{impequation}) we have  to  show that the ``number'' of paths
 $A\rightarrow B$ that flip and  the ``number" of paths $\bar{A}\rightarrow B$
 converge asymptomatically.This is done in $(ii)$.
 In the subsequent analysis we will
 consider a $d$ dimensional lattice - lattice spacing being infinitesimally
 small- instead of continuum  for the sake of notational convenience only.
 The reason for (\ref{impequation})
 follows. \\
{\bf{i)}}     An initial configuration at $A$ of Fig 1
  given by $X_{AB}=\{...\alpha_1,\alpha,\alpha_2,...\}$, is considered,
where $\alpha_1$,$\alpha_2$ ...  are the initial values of $\phi$ at coordinates
 $x\neq x_o$ . The corresponding path
 takes initial $\phi(x_o)=\alpha$
 to $B$ , then   it may be concluded from (2) that $X_{\bar{A}B}=
\{...f\alpha_1,-\alpha,f\alpha_2,...\}$  ($f=\frac{\beta+{(4\pi t)}^{-d/2}\alpha}
{\beta-{(4\pi t)}^{-d/2}\alpha}$) is a configuration at  $\bar{A}$ which takes  initial $\phi(x_o)=-\alpha$
 to $B$. Hence there is a one to one mapping of  paths from $A\rightarrow B$
to those  from $\bar{A}\rightarrow B$ . It may be underlined here that
$f\rightarrow 1$ as $t\rightarrow \infty$. This implies that the probability
 of the two paths approach each other asymptotically.\\
 {\bf{ii)}} In this part we will address the fact that in the asymptotically large
 time limit it is a very good approximation  to say that there
 is a one to one
 correspondence between the paths from $A$ that flip to those from
  $\bar{A} \rightarrow  B$.This   may be used  as it is a  controlled  approximation for  it improves  with increasing $t$.
 In order to see this point let us consider a point $C$
$({t_2},\beta)$(not shown in the $Fig$ $1$) where
${t_2}>t_1$. Let $Y_{AB}=\{...\gamma_1,\alpha,\gamma_2,...\}$ be
 the initial configuration corresponding to path $ADB$ ( the
 path in blue in Fig 1) where $\gamma_1$,$\gamma_2$ ...  are the initial values of $\phi$ at coordinates
 $x\neq x_o$.
 This path   crosses zero while reaching $B$. It can
 be  shown that
 $Y_{AC}=
\{...f_1\gamma_1,\alpha,f_1\gamma_2,...\}$
($f_1=(\frac{t_2}{t_1})^{d/2}
\frac{\beta -{(4\pi t_2)}^{-d/2}\alpha)}
{\beta -{(4\pi t_1)}^{-d/2}\alpha}$)  is the corresponding initial configuration 
 for a path  $A\rightarrow C$. The exact expression for
 $f_1$ contains a  coordinate dependent term whose leading order
 behavior for large $t$ is $1$.
 Since ${t_2}>t_1$,
 we have  $f_1>1$ for sufficiently
 large $t_1$. Let the time coordinate at $D$ be $t_D$, then
$\phi (x_o,t_D)=0$ for the path $ADB$. Then one may arrive from
(\ref{third}) that   $\phi (x_o,t_D)<0$ for the initial configuration $Y_{AC}$.Hence, one can conclude that the path corresponding to
$Y_{AC}$ must have flipped at an earlier time than $t_D$.
Therefore, if a path from $A\rightarrow B$ flips, the corresponding
 path from $A\rightarrow C$ flips at an earlier time.
 Since ${t_2}>t_1$, the `number' of paths flipping while
 going from  $A\rightarrow C$  is
 more than  those from $A\rightarrow B$. Thus the `number' of paths from
  $A\rightarrow B$ that flip  is a  fraction $f_2$ of those from
 ${\bar{A}}\rightarrow B$
 where $f_2=1-O({t_1}^{-a})$ for large $t_1$,$a$ being some positive number. \\

 On account of (i), (ii) we say that the probability of the paths( like
 ADB in Fig 1) that  flip  while reaching $B$ in the large time limit is
 given by $ P\big(\beta|-\alpha\big)h_{correction}$, where
 $h_{correction}= 1 + O(t^{-b})$ ,  $b=1$, being Taylor 
 expansion in $t^{-1}$.  In principle, the coefficient of $t^{-b}$ may be a function of 
 $\beta$. When integrating over $\beta$ - as will be done later- 
 the contribution to the integral comes from the vicinity of $\beta = -\mu \sim 
 t^{-d/2}$ which is vanishingly small in the asymptotic limit. Also, 
 $d\beta$ $\sim$ $\sigma$ $\sim$ $t^{-d/4}$. The coefficient
 is  Taylor expanded about $\beta = 0$ and only the    
 zeroth order term or the term 
 independent of $\beta$ is retained. So the probability of $\phi(x_o)$ not changing
 sign when reaching the neighborhood($d\beta$) of $B$ is,  for asymtotically large time,

\ber\label{asymtote}
 \lim_{t\to\infty}
\lefteqn{ \big[P\big(\beta|\alpha\big)d\beta- P\big(\beta|
-\alpha\big)h_{correction}d\beta\big] }\nonumber\\
&& =P\big(\beta|\alpha\big)d\beta- P\big(\beta|
-\alpha\big)\big( 1 + O(t^{-1})\big) d\beta
\eer

 This leads us to  (\ref{impequation}). The final $\beta$ may have any value
 as long as it remains positive. The probability of $\phi(x_o)$
starting from  $\alpha$
 and reaching a final  positive value without ever changing sign is

\beq\label{impequation2}
\mathbb{P}^{+}\big(\alpha\big) =\int^\infty_0 \,d\beta P^{+}
\big(\beta|\alpha\big)
\eeq

 We would now calculate (\ref{impequation2}) for asymptotically large value of t.Under the
 circumstances the second term on the R.H.S of (\ref{ninth}) can be neglected.
 Further  $\frac{{\mu}^{2}}{{\sigma}^{2}}\sim \alpha t^{-d/2}$. Hence for
 $\alpha\ll t^{d/2}$, ${\mu}^{2}/{\sigma}^{2}\ll 1$. The expression
 (\ref{impequation2}) is evaluated using the identity\cite{gradshteyn2014table}
\beq
\fbox{$ \displaystyle \int^\infty_0 \,dx\exp (\frac{-x^{2}}
{4\beta}  - \gamma x) =
\sqrt{\pi\beta}\exp(\beta{\gamma}^{2})\big[1 - erf(\gamma\sqrt{\beta})\big]$}
\eeq

Evaluation leads to a sum of two terms - one is proportional to 
 $t^{-d/4}$ and the other is proportional to $t^{-1}$. Hence we obtain
\beq\label{impequation3}
\mathbb{P}^{+}\big(\alpha\big) \sim \alpha t^{-d/4}
\eeq
 for $d\leq 4$.
 In arriving at the   above result the asymptotic expansion of
 `error function' $erf$ has been used for small argument.
  Finally, the  expression for $\mathcal{P}^{+} (t)$ is obtained by integrating $\alpha$ over a Gaussian distribution.
\beq\label{impequation4}
\mathcal{P}^{+} (t) = \int^\infty_0 \,d\alpha \mathbb{P}^{+}\big(\alpha\big) Q\big(\alpha\big)
\eeq
 where $ Q(\alpha)$ is the Gaussian distribution for initial $\phi(x_o,o)=\alpha$
 with variance $ k$ as mentioned at  the beginning.
 If $k<< t^{d/2}$, it may be concluded
 from (\ref{impequation3}) and (\ref{impequation4}) that $\mathcal{P}^{+}(t) \sim t^{-d/4}$  or $t^{-1}$ depending on whether $d\leq 4$ or not.
 This gives $\theta_o=d/4$  or $1$ \\

\section{Result and Conclusion:}
 In the previous section exact calculation has been carried out to determine
 the probability  $\mathcal{P}^{+}(t)$ of the  sign of the field
$\phi$ remaining positive through
 out  a asymtotically large time $t$. The probability is  $\mathcal{P}^{+}(t) \sim t^{-d/4}$. Hence the persistence exponent  is $\theta_o=d/4$ valid for any
 arbitrary integer dimension $d\leq 4$. The exponents for $d = 1$, $2$ , $3$ are
 $0.25$, $0.50$, $0.75$ respectively. \\
 The result may be experimentally  verified for a system initially
 at thermal equilibrium defined by a temperature T. The equilibrium is 
 then disturbed  in a suitable manner. The time evolution of the coarse
 grained temperature at any point satisfies the simple diffusion equation,
hence this time evolution can be  studied to find the persistence
 exponent. \\
The answer for the exponent $\theta_o$ obtained in this paper  is 
in disagreement  with all the papers cited  in the beginning. The first results for the persistence
 exponent in the case of the  diffusion problem were published in
 \cite{majumdar1996nontrivial},\cite{derrida1996persistent} back to back. The
 papers used a two time correlation function and explicitly applied the
approximation (IIA), Independent Interval Approximation{\color{red},}
to get to the answer. Application of the two time correlation function  is not suitable here
 and so is IIA which   is a Markovian approximation . Further , the papers use Monte Carlo
 simulation to confirm the result. Monte Carlo method appears to be
 unsuitable for this problem. Hence all the papers that reproduce the
 results  of  \cite{majumdar1996nontrivial,derrida1996persistent}
 are not expected to give the correct answer. In \cite{ehrhardt2002series} the  authors have defined a correlation function $C(T)$- just like
  \cite{majumdar1996nontrivial,derrida1996persistent}- to carry the
 calculation forward.
Let us also consider \cite{poplavskyi2018exact} where
 the authors have used Kac Polynomials \cite{kac1943average}
 to obtain  the `exact exponent' in 2d. 
The answer obtained agrees perfectly
 with \cite{majumdar1996nontrivial,derrida1996persistent}.
In the course of the calculation they have used  that the zero crossing
 property is    governed by the covariance $c(T)=sech(\frac{T}{2})$
 \cite{poplavskyi2018exact} of the stationary Gaussian process i.e the
 diffusion equation with time redefined. Similarly in \cite{hilhorst2000persistence},
 correlator in time $F_{\epsilon}(\tau - \tau ')$ has been used in the
 calculation.The
 point is that the covariance/correlator/correlation function is a
 misleading  quantity  for the problem
  for reasons mentioned below. The model presented in the
 paper  has randomness only in the intial condition. Once the system
 starts evolving there is no further randomness. It evolves in accordance
 with the kernel in  (\ref{third}). It is encoded in  the initial
 condition when and where the $\phi$ will flip.The
 probability of each path is uniquely determined by the probability
 of initial condition, hence the problem
 with covariance/correlator.The covariance function  imposes
 stochasticity on the present problem   throughout the entire time evolution.
 We now have a different model with the same correlation function but no
 unique dependence of the probability of the path on initial condition.  It also
 makes the problem  Markovian. Hence , all the previous results
  are in perfect agreement though the calculated exponent
will be different from the actual value. The value of the
 exponent does not depend on  only correlation function but
 it depends on other details of the model too.  Further , there even
 appears to be experimental proof \cite{wong2001measurement} for the
results of
\cite{majumdar1996nontrivial}, \cite{derrida1996persistent}. The experimental
 setup of \cite{wong2001measurement}
does not  represent the diffusion model described in this paper. The setup
  satisfies the  approximations of the previous
 papers  and hence the agreement with
 the  their result. \\

\section*{Acknowledgements}

  The author would like to thank CSIR, India  for
 Fellowship during the course of the work (2004) at IACS, India \\

\bibliographystyle{unsrt}

\bibliography{references}

\end{document}